%% file: RMF-OMEG.tex
\documentclass[final,5p,times,twocolumn]{elsarticle}
\usepackage{amsmath}
\usepackage{bm}
\usepackage{booktabs}
\usepackage{braket}
\usepackage{caption}
\usepackage{color}
\usepackage{graphicx}
\usepackage[colorlinks]{hyperref}

\journal{}

\begin{document}

\begin{frontmatter}
  \title{Can the PREX-2 and CREX results be understood by relativistic mean-field \\ models with the astrophysical constraints?}
  \author[1]{Tsuyoshi Miyatsu}
  \address[1]{Department of Physics and OMEG Institute, Soongsil University, Seoul 06978, Republic of Korea}
  \ead{tsuyoshi.miyatsu@ssu.ac.kr}
  \author[1]{Myung-Ki Cheoun}
  \ead{cheoun@ssu.ac.kr}
  \author[2]{Kyungsik Kim}
  \address[2]{School of Liberal Arts and Sciences, Korea Aerospace University, Goyang 10540, Republic of Korea}
  \ead{kyungsik@kau.ac.kr}
  \author[3]{Koichi Saito}
  \address[3]{Department of Physics and Astronomy, Faculty of Science and Technology, Tokyo University of Science, Noda 278-8510, Japan}
  \ead{koichi.saito@rs.tus.ac.jp}

  \begin{abstract}
    We construct new effective interactions using the relativistic mean-field models with the isoscalar- and isovector-meson mixing, $\sigma^{2}\bm{\delta}^{2}$ and $\omega_{\mu}\omega^{\mu}\bm{\rho}_{\nu}\bm{\rho}^{\nu}$.
    Taking into account the particle flow data in heavy-ion collisions, the observed mass of PSR J0740$+$6620, and the tidal deformability of a neutron star from binary merger events, GW170817 and GW190814, we study the ground-state properties of finite, closed-shell nuclei, and try to explain the recent results from the PREX-2 and CREX experiments.
    It is found that the $\sigma$--$\delta$ mixing is very powerful to understand the terrestrial experiments and astrophysical observations of neutron stars self-consistently.
    We can predict the large neutron skin thickness of $^{208}$Pb, $R_{\rm skin}^{208}=0.243$~fm, using the slope parameter of nuclear symmetry energy, $L=70$~MeV, which is consistent with the PREX-2 result.
    However, to explain the CREX data, it is preferable to adopt the small value of $L=20$~MeV.
    It is very difficult to understand the PREX-2 and CREX results simultaneously within relativistic mean-field models.
  \end{abstract}


\end{frontmatter}

\section{Introduction} \label{sec:introduction}

The accurate measurement of neutron skin thickness of $^{208}$Pb, $R_{\rm skin}^{208}$, from the parity-violating electron scattering by the PREX Collaboration has revealed a serious discrepancy between the measured $R_{\rm skin}^{208}$ and that expected theoretically~\cite{PREX:2021umo}.
In order to explain the result from the PREX-2 experiment, \citet{Reed:2021nqk} have proposed the large value of the slope parameter of nuclear symmetry energy, $L=106\pm37$~MeV, by exploiting the strong correlation between $R_{\rm skin}^{208}$ and $L$.
In contrast, using modern relativistic and nonrelativistic energy density functionals, \citet{Reinhard:2021utv} have predicted the relatively small value, $L=54\pm8$~MeV, by the assessment of theoretical uncertainty on the parity-violating asymmetry, $A_{\rm PV}$, in $^{208}$Pb.
Furthermore, the precise measurement of neutron skin thickness of $^{48}$Ca, $R_{\rm skin}^{48}$, has been performed by the CREX Collaboration~\cite{CREX:2022kgg}.
Then, it turns out that it is very important to determine the value of $L$ accurately~\cite{Piekarewicz:2021jte,SensuremathpiRIT:2021xau,Reinhard:2022inh}.

Relativistic mean-field (RMF) calculations have achieved great success in understanding of the properties of nuclear matter and neutron stars as well as finite nuclei~\cite{Serot:1984ey,Glendenning:1991es}.
In our previous study~\cite{Miyatsu:2022wuy}, we have developed the RMF model by introducing the isoscalar- and isovector-meson mixing, $\sigma^{2}\bm{\delta}^{2}$ and $\omega_{\mu}\omega^{\mu}\bm{\rho}_{\nu}\bm{\rho}^{\nu}$.
It has been found that the quartic interaction of the scalar mesons has large influence on the radius and tidal deformability of a neutron star.
In particular, the strong $\sigma$--$\delta$ mixing could reproduce the dimensionless tidal deformability of a canonical $1.4$~$M_{\odot}$ neutron star, $\Lambda_{1.4}$, observed by the gravitational-wave (GW) signals from binary neutron star merger, GW170817~\cite{LIGOScientific:2017vwq}.
Furthermore, the effect of $\delta$ meson on the characteristics of finite nuclei has been recently studied in detail~\cite{Li:2022okx,Thakur:2022dxb}.

In the present study, focusing on the PREX-2 and CREX experiments, we construct new effective interactions using the RMF model under the constraints from the terrestrial experiments and astrophysical observations of neutron stars.
It is necessary that the resulting nuclear equation of state (EoS) follows the conditions:
(1) the EoSs for symmetric nuclear matter and pure neutron matter satisfy the particle flow data in heavy-ion collisions (HICs)~\cite{Danielewicz:2002pu,Fuchs:2005zg,Lynch:2009vc},
(2) the EoS for neutron stars attains to the observed mass of PSR J0740$+$6620 ($M=2.072^{+0.067}_{-0.066}$~$M_{\odot}$)~\cite{NANOGrav:2019jur,Fonseca:2021wxt,Riley:2021pdl}, and
(3) the EoS for neutron stars explains the tidal deformability from both binary merger events, GW170817 ($\Lambda_{1.4}=190^{+390}_{-120}$)~\cite{LIGOScientific:2018cki,LIGOScientific:2018hze} and GW190814 ($\Lambda_{1.4}=616_{-158}^{+273}$)~\cite{LIGOScientific:2020zkf}.
Under these constraints, we examine the effects of the $\delta$-nucleon coupling and $\sigma$--$\delta$ mixing on the ground-state properties of finite nuclei, and consider the PREX-2 and CREX results.


\section{Formalism} \label{sec:formalism}

We employ the recently updated Lagrangian density including the isoscalar ($\sigma$ and $\omega^{\mu}$) and isovector ($\bm{\delta}$ and $\bm{\rho}^{\mu}$) mesons as well as nucleons ($N=p,n$)~\cite{Miyatsu:2022wuy}.
The interacting Lagrangian density is then given by
\begin{align}
  \mathcal{L}_{\rm int}
  & = \sum_{N}\bar{\psi}_{N}\bigl[g_{\sigma}\sigma
    - g_{\omega}\gamma_{\mu}\omega^{\mu}
    + g_{\delta}\bm{\delta}\cdot\bm{\tau}_{N}
    \nonumber \\
  & - g_{\rho}\gamma_{\mu}\bm{\rho}^{\mu}\cdot\bm{\tau}_{N}\bigr]\psi_{N}
    - U_{\rm NL}(\sigma,\omega,\bm{\delta},\bm{\rho}),
    \label{Lint}
\end{align}
where $\psi_{N}$ is the nucleon field and $\bm{\tau}_{N}$ is its isospin matrix.
The meson-nucleon coupling constants are respectively denoted by $g_{\sigma}$, $g_{\omega}$, $g_{\delta}$, and $g_{\rho}$.
The nonlinear potential is here supplemented as
\begin{align}
  U_{\rm NL}(\sigma,\omega,\bm{\delta},\bm{\rho})
  & = \frac{1}{3}g_{2}\sigma^{3} + \frac{1}{4}g_{3}\sigma^{4} - \frac{1}{4}c_{3}\left(\omega_{\mu}\omega^{\mu}\right)^{2}
    \nonumber \\
  & - \Lambda_{\sigma\delta}\sigma^{2}\bm{\delta}^{2}
    - \Lambda_{\omega\rho}\left(\omega_{\mu}\omega^{\mu}\right)\left(\bm{\rho}_{\nu}\cdot\bm{\rho}^{\nu}\right),
    \label{eq:NLpot}
\end{align}
with three coupling constants, $g_{2}$, $g_{3}$, and $c_{3}$, and two mixing parameters, $\Lambda_{\sigma\delta}$ and $\Lambda_{\omega\rho}$~\cite{Boguta:1977xi,Sugahara:1993wz,Todd-Rutel:2005yzo,Miyatsu:2013yta,Zabari:2018tjk}.
For describing the characteristics of finite nuclei, the Coulomb interaction for the photon field, $A^{\mu}$, is also taken into account~\cite{Serot:1984ey,Ring:1996qi}.

In RMF approximation, the meson and photon fields are replaced by the mean-field values: $\bar{\sigma}$, $\bar{\omega}$, $\bar{\delta}$, $\bar{\rho}$, and $\bar{A}$.
Then, the effective nucleon mass in matter is simply expressed as
\begin{equation}
  M_{p \choose n}^{\ast}(\bar{\sigma},\bar{\delta}) = M_{N} - g_{\sigma}\bar{\sigma} \mp g_{\delta}\bar{\delta},
\end{equation}
where $M_{N}$~($=939$ MeV) is the nucleon mass in free space.
If we restrict consideration to spherical nuclei, the equations of motion for the nucleon, meson, and photon fields are given by
\begin{align}
  \Biggl[-i\bm{\alpha}\cdot\bm{\nabla}+\beta
  & M_{p \choose n}^{\ast} + g_{\omega}\bar{\omega}
    \nonumber \\
  & \pm g_{\rho}\bar{\rho} + e\frac{1\pm1}{2}\bar{A}
    \Biggr]\,\psi_{p \choose n} = \varepsilon_{\alpha{p \choose n}}\psi_{p \choose n},
    \label{eq:Dirac} \\
  \left(-\bm{\nabla}^{2}+m_{\sigma}^{2}\right)\bar{\sigma}
  & = g_{\sigma}\left(\rho_{p}^{s}+\rho_{n}^{s}\right)
    \nonumber \\
  & - g_{2}\bar{\sigma}^{2}-g_{3}\bar{\sigma}^{3}+2\Lambda_{\sigma\delta}\bar{\sigma}\bar{\delta}^{2},
    \label{eq:sigma} \\
  \left(-\bm{\nabla}^{2}+m_{\omega}^{2}\right)\bar{\omega}
  & = g_{\omega}\left(\rho_{p}+\rho_{n}\right)-c_{3}\bar{\omega}^{3}-2\Lambda_{\omega\rho}\bar{\omega}\bar{\rho}^{2},
    \label{eq:omega} \\
  \left(-\bm{\nabla}^{2}+m_{\delta}^{2}\right)\bar{\delta}
  & = g_{\delta}\left(\rho_{p}^{s}-\rho_{n}^{s}\right)+2\Lambda_{\sigma\delta}\bar{\sigma}^{2}\bar{\delta},
    \label{eq:delta} \\
  \left(-\bm{\nabla}^{2}+m_{\rho}^{2}\right)\bar{\rho}
  & = g_{\rho}\left(\rho_{p}-\rho_{n}\right)-2\Lambda_{\omega\rho}\bar{\omega}^{2}\bar{\rho},
    \label{eq:rho} \\
  -\bm{\nabla}^{2}\bar{A}
  & = e\rho_{p},
    \label{eq:photon}
\end{align}
with $\varepsilon_{\alpha N}$ being the nucleon single-particle energy, and $\rho_{N}^{s}$ ($\rho_{N}$) being the scalar (baryon) density for $N$~$(=p,n)$.
The total energy of the system is then given by
\begin{align}
  E
  & = \sum_{N=p,n}\sum_{\alpha}^{\rm occ}\left(2j_{\alpha}+1\right)\varepsilon_{\alpha N}
    \nonumber \\
  & + \frac{1}{2}\int d\bm{r} \,\Biggl[g_{\sigma}\left(\rho_{p}^{s}+\rho_{n}^{s}\right)\bar{\sigma}
    - g_{\omega}\left(\rho_{p}+\rho_{n}\right)\bar{\omega}
    \nonumber \\
  & + g_{\delta}\left(\rho_{p}^{s}-\rho_{n}^{s}\right)\bar{\delta}
    - g_{\rho}\left(\rho_{p}-\rho_{n}\right)\bar{\rho}-e\rho_{p}\bar{A}
    \nonumber \\
  & - \frac{1}{3}g_{2}\bar{\sigma}^{3}
    - \frac{1}{2}g_{3}\bar{\sigma}^{4}+\frac{1}{2}c_{3}\bar{\omega}^{4}
    + 2\Lambda_{\sigma\delta}\bar{\sigma}^{2}\bar{\delta}^{2}+2\Lambda_{\omega\rho}\bar{\omega}^{2}\bar{\rho}^{2}\Biggr]\,,
\end{align}
where the sum $\alpha$ runs over the occupied states of $\varepsilon_{\alpha N}$ with the degeneracy $\left(2j_{\alpha}+1\right)$.

\section{Results and discussions} \label{sec:results}

We here propose three effective interactions labeled as OMEG1, OMEG2, and OMEG3.
For the sake of comparison, the two well-known interactions, IOPB-I~\cite{Kumar:2017wqp} and FSUGarnet~\cite{Chen:2014mza}, are also used in the present calculation, which are consistent with the flow data in HICs and the maximum-mass condition of a neutron star.
However, they do not support the severe constraint on $\Lambda_{1.4}$ from GW170817.

\input{./Tables/Table-CCs.tex}
The model parameters for the OMEG family, IOPB-I, and FSUGarnet are given in Table~\ref{tab:CCs}.
The model optimization in the OMEG family is performed so as to fit the experimental data for binding energy per nucleon, $B/A$, of several closed-shell nuclei, and charge radius, $R_{\rm ch}$, of $^{40}$Ca~\cite{Serot:1984ey}.
\input{./Tables/Table-finite.tex}
The theoretical predictions are given in Table~\ref{tab:finite}.
We here consider the zero-point energy correction taken from the conventional Skyrme Hartree--Fock calculations~\cite{Reinhard:1986qq,Sugahara:1993wz}.
The $R_{\rm ch}$ is defined as
\begin{equation}
  R_{\rm ch}=\sqrt{R_{p}^{2}+(0.8783)^{2}},
\end{equation}
with $R_{p}$ being the point proton radius~\cite{Chen:2014sca}.
In addition, the root relative squared errors (RRSEs),
\begin{equation}
  \Delta_{\rm RRSE}
  = \sqrt{\frac{1}{n}\sum_{i=1}^{n}\left(\frac{X_{i}^{\rm expt.}-X_{i}^{\rm calcn.}}{X_{i}^{\rm expt.}}\right)^{2}},
\end{equation}
for $X=B/A$ and $R_{\rm ch}$ are given in the last column of Table~\ref{tab:finite}.
We note that the OMEG family, as well as the IOPB-I and FSUGarnet, can reproduce the properties of finite nuclei very well.

\begin{figure}[t!]
  \includegraphics[width=8.5cm,keepaspectratio,clip]{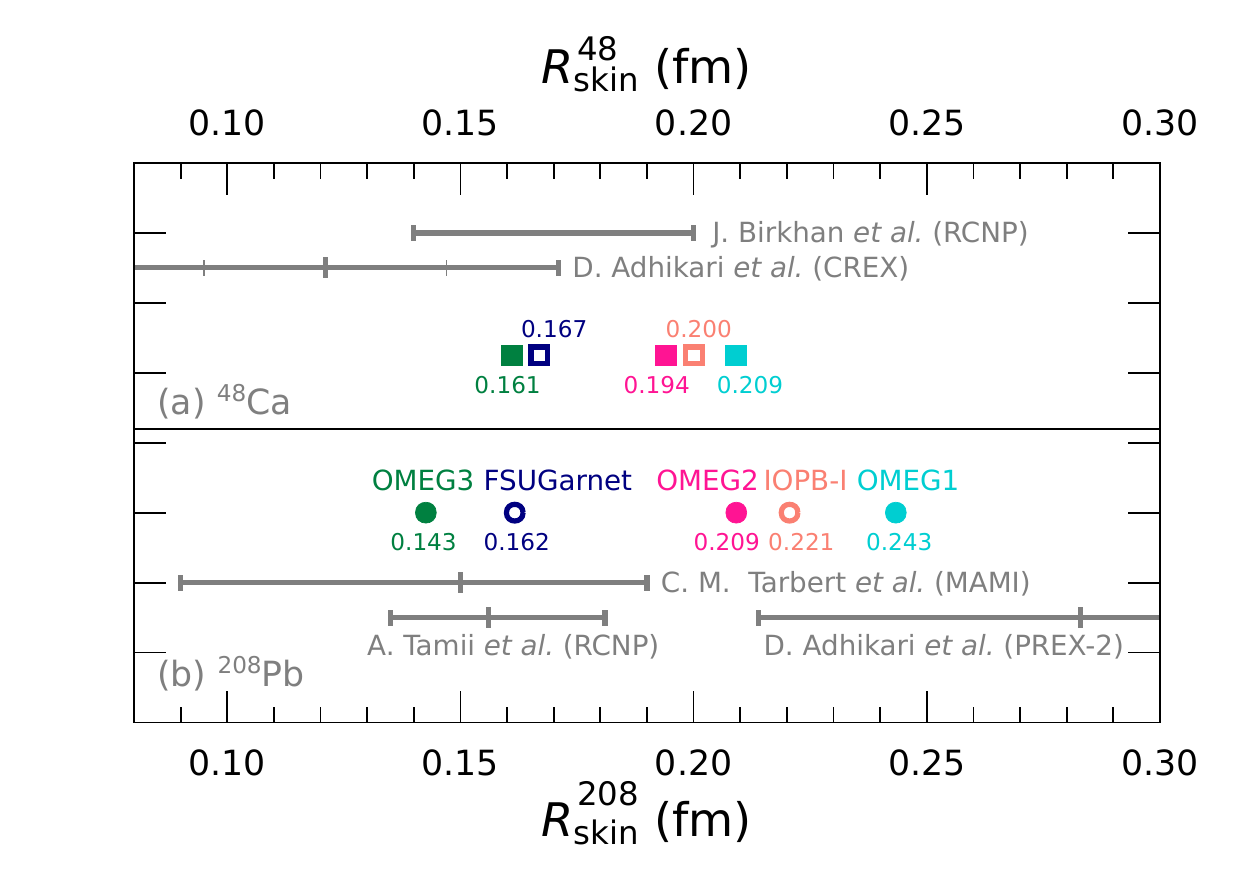}%
  \caption{\label{fig:Rskin}
    Neutron skin thickness of (a) $^{48}$Ca, $R_{\rm skin}^{48}$, and of (b) $^{208}$Pb, $R_{\rm skin}^{208}$.}
\end{figure}
In Fig.~\ref{fig:Rskin}, we summarize the theoretical results of $R_{\rm skin}^{48}$ and $R_{\rm skin}^{208}$, compared with the experimental data: the electric dipole polarizability of $^{48}$Ca (RCNP; $R_{\rm skin}^{48}=0.14$--$0.20$~fm)~\cite{Birkhan:2016qkr}, the complete electric dipole response on $^{208}$Pb (RCNP; $R_{\rm skin}^{208}=0.156_{-0.021}^{+0.025}$~fm)~\cite{Tamii:2011pv}, the coherent pion photoproduction cross sections measurement of $^{208}$Pb (MAMI; $R_{\rm skin}^{208}=0.15\pm0.03(\mathrm{stat.})_{-0.03}^{+0.01}(\mathrm{sys.})$~fm)~\cite{Tarbert:2013jze}, and the parity-violating electron scattering off $^{48}$Ca (CREX; $R_{\rm skin}^{48}=0.121\pm0.026(\mathrm{exp.})\pm0.024(\mathrm{model})$~fm)~\cite{CREX:2022kgg} and off $^{208}$Pb (PREX-2; $R_{\rm skin}^{208}=0.283\pm0.071$~fm)~\cite{PREX:2021umo}.

The OMEG1 gives the largest value, $R_{\rm skin}^{208}=0.243$~fm, which meets the PREX-2 result.
The OMEG2 is selected so as to match the predicted result, $R_{\rm skin}^{208}=0.19\pm0.02$~fm, by the assessment of the theoretical uncertainty on $A_{\rm PV}$ in $^{208}$Pb~\cite{Reinhard:2021utv}.
Meanwhile, the OMEG3 satisfies all the experiments except the PREX-2 data.
Therefore, as mentioned by \citet{Reinhard:2022inh}, it is very hard to understand the PREX-2 and CREX results simultaneously even if the $\delta$--$N$ coupling and $\sigma$--$\delta$ mixing are taken into consideration in the RMF model.

\input{./Tables/Table-matter.tex}

\begin{figure}[t!]
  \includegraphics[width=8.5cm,keepaspectratio,clip]{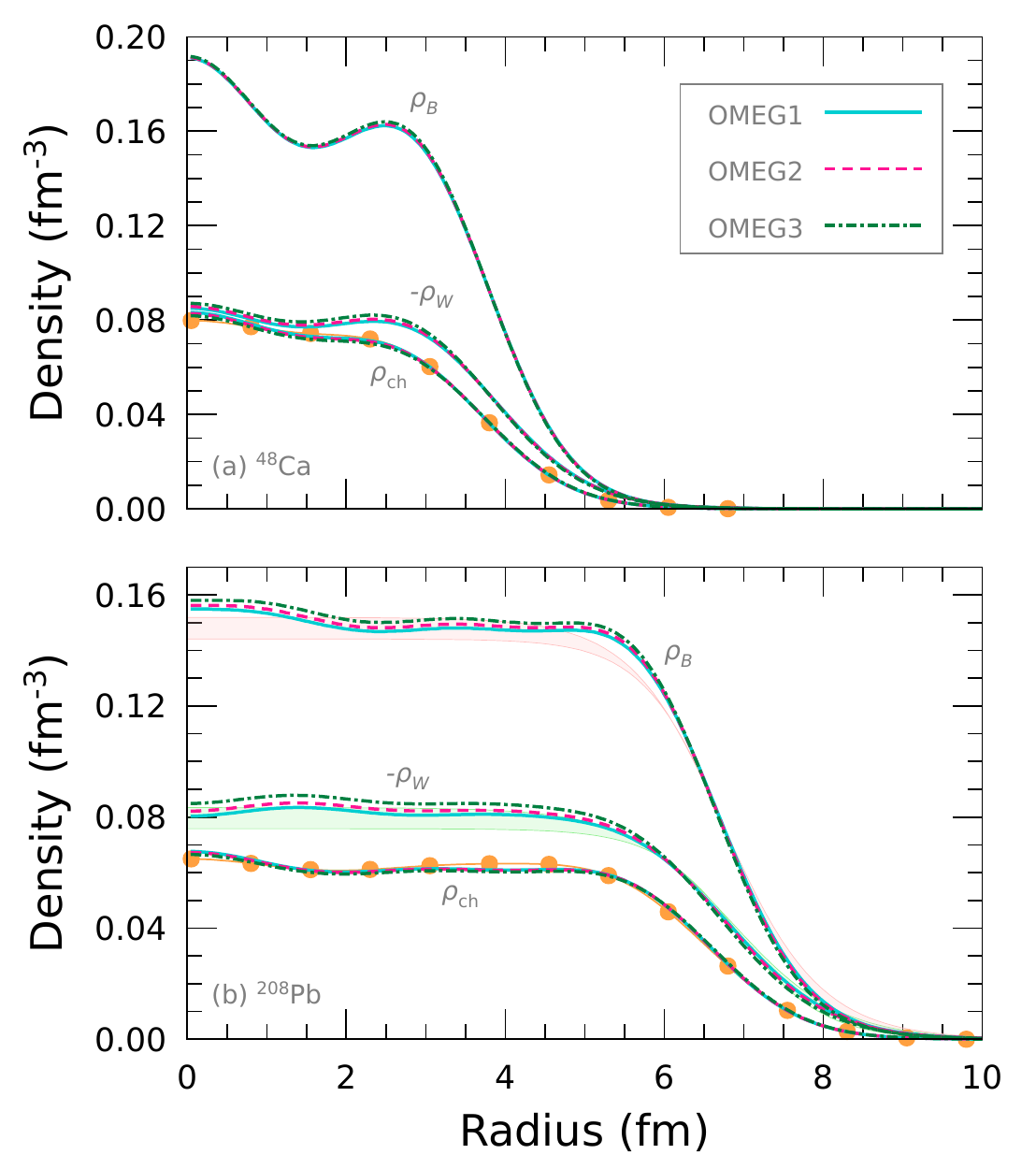}%
  \caption{\label{fig:Nuclei}
    Baryon, charge, and weak charge densities, $\rho_{B}$, $\rho_{\rm ch}$, and $\rho_{W}$, by the OMEG family for (a) $^{48}$Ca and for (b) $^{208}$Pb.
    The shaded bands represent the experimental data of $-\rho_{W}$ and $\rho_{B}$ by the PREX Collaboration~\cite{PREX:2021umo}, and the dots are the result of $\rho_{\rm ch}$ measured from the elastic electron scattering~\cite{DeVries:1987atn}.}
\end{figure}
The density profiles in $^{48}$Ca and $^{208}$Pb are displayed in Fig.~\ref{fig:Nuclei}.
We here present the baryon, charge, and weak charge densities, $\rho_{B}$~$(=\rho_{p}+\rho_{n})$, $\rho_{\rm ch}$, and $\rho_{W}$, with the experimental results~\cite{PREX:2021umo,DeVries:1987atn}.
The $\rho_{W}$ is expressed as
\begin{equation}
  \rho_{W}\left(\bm{r}\right)
  \simeq Q_{p}\rho_{\rm ch}\left(\bm{r}\right)
  +      Q_{n}\int d\bm{r}^{\prime} \, G_{p}^{E}\left(\left|\bm{r}-\bm{r}^{\prime}\right|\right)\rho_{n}\left(\bm{r}\right),
\end{equation}
with $Q_{p(n)}$ being the proton (neutron) weak charge and $G_{p}^{E}$ being the proton electric form factor~\cite{Horowitz:1999fk,Horowitz:2012tj}.
The OMEG members are calibrated so as to reproduce $-\rho_{W}$ and $\rho_{B}$ in $^{208}$Pb by the PREX-2 experiment.

We present, in Table~\ref{tab:matter}, several bulk properties of symmetric nuclear matter at the saturation density, $\rho_{0}$.
Here, the nuclear incompressibility in the OMEG family is determined as $K_{0}=256$~MeV.
This value is consistent with the comprehensive, relativistic reanalysis of recent data on giant monopole resonances~\cite{Stone:2014wza}, which suggests $250<K_{0}\mathrm{\,(MeV)}<315$.
We note that the OMEG family has a relatively larger $M_{N}^{\ast}$ than those for the IOPB-I and FSUGarnet, because $M_{N}^{\ast}$ is very sensitive to the stringent constraint on $\Lambda_{1.4}$ observed from GW170817~\cite{Hornick:2018kfi,Choi:2020eun}.

As $L$ is strongly correlated with $R_{\rm skin}^{48}$ and $R_{\rm skin}^{208}$, the OMEG members resultantly cover a wide range of $20\leq L\mathrm{\,(MeV)}\leq70$.
It is noticeable that the curvature parameters, $K_{\rm sym}$, for the OMEG1 and OMEG2 are smaller than that by the theoretical analyses based on neutron-star observables, $K_{\rm sym}\approx-107\pm88$~MeV~\cite{Li:2021thg}.
Correspondingly, the neutron matter incompressibility, $K_{N}$, in the OMEG1 and OMEG2 becomes much smaller than that predicted by the chiral effective field theory, $K_{N}=152.2\pm38.1$~MeV~\cite{Lattimer:2023rpe}.
Furthermore, the asymmetry term of nuclear incompressibility, $K_{\rm sat,2}$, for the OMEG1 satisfies the experimental constraints from isoscalar giant monopole resonances in the Sn and Cd isotopes, $K_{\rm sat,2}=-550\pm100$ MeV~\cite{Li:2010kfa} and $K_{\rm sat,2}=-555\pm75$ MeV~\cite{Patel:2012zd}, respectively.

\begin{figure}[t!]
  \includegraphics[width=8.5cm,keepaspectratio,clip]{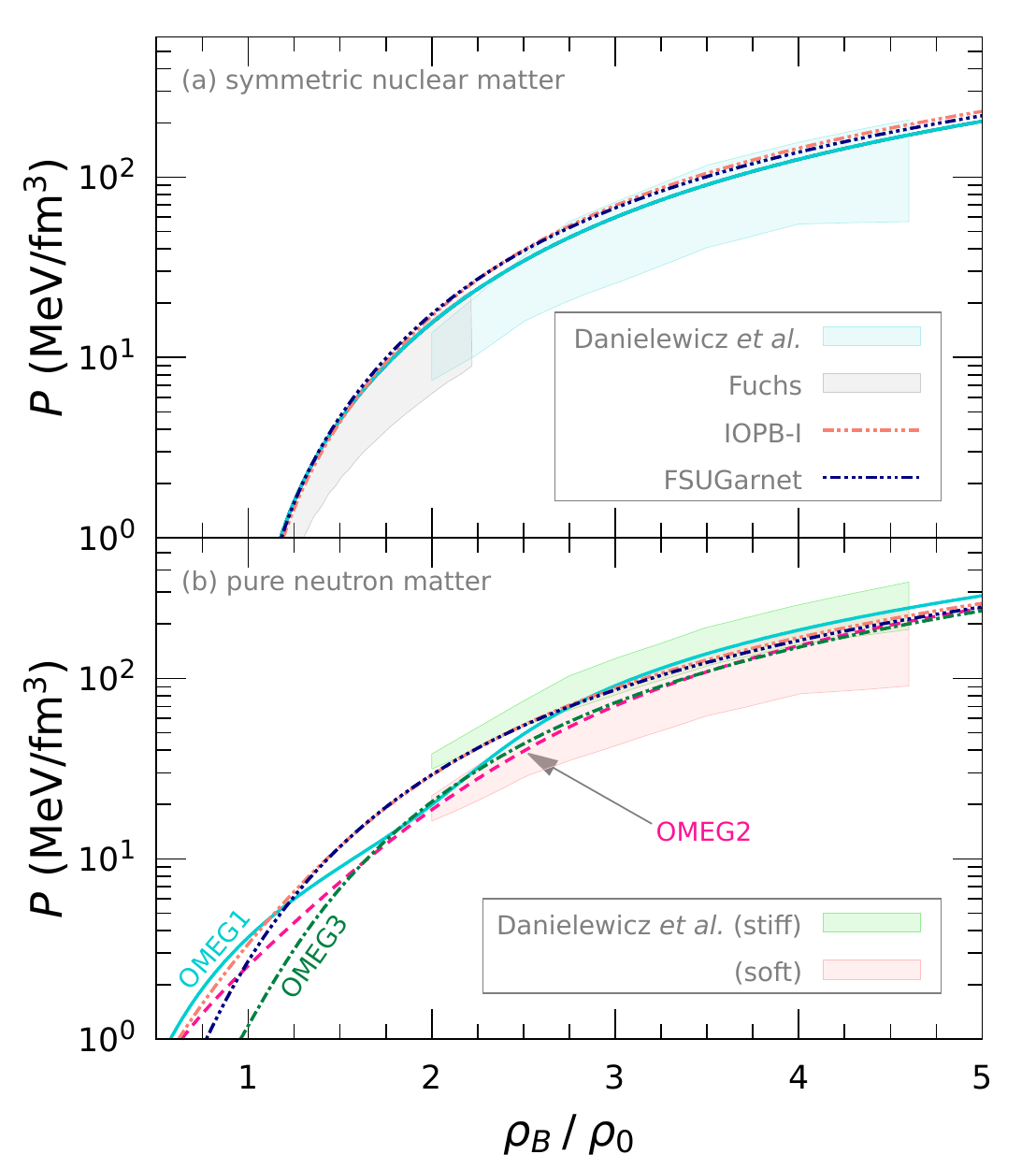}%
  \caption{\label{fig:Pressure}
    EoS---pressure, $P$, as a function of $\rho_{B}/\rho_{0}$---for (a) symmetric nuclear matter and for (b) pure neutron matter.}
\end{figure}
The EoSs for symmetric nuclear matter and pure neutron matter are depicted in Fig.~\ref{fig:Pressure}.
In both cases, we show the constraints from elliptical flow data~\cite{Danielewicz:2002pu} and kaon production data~\cite{Fuchs:2005zg,Lynch:2009vc} in HICs.
In symmetric nuclear matter, pressure for the OMEG family can satisfy the constraints from HICs by means of the $\omega$ self-interaction, which is similar to the cases of the IOPB-I and FSUGarnet.
It is found that, in pure neutron matter, pressure around $2\rho_{0}$ for the OMEG family is softer than those for the IOPB-I and FSUGarnet.
At high densities, the OMEG1 shows the harder EoS than any others because of the small $\omega$--$\rho$ mixing, as shown in Table~\ref{tab:CCs}.

\begin{figure}[t!]
  \includegraphics[width=8.5cm,keepaspectratio,clip]{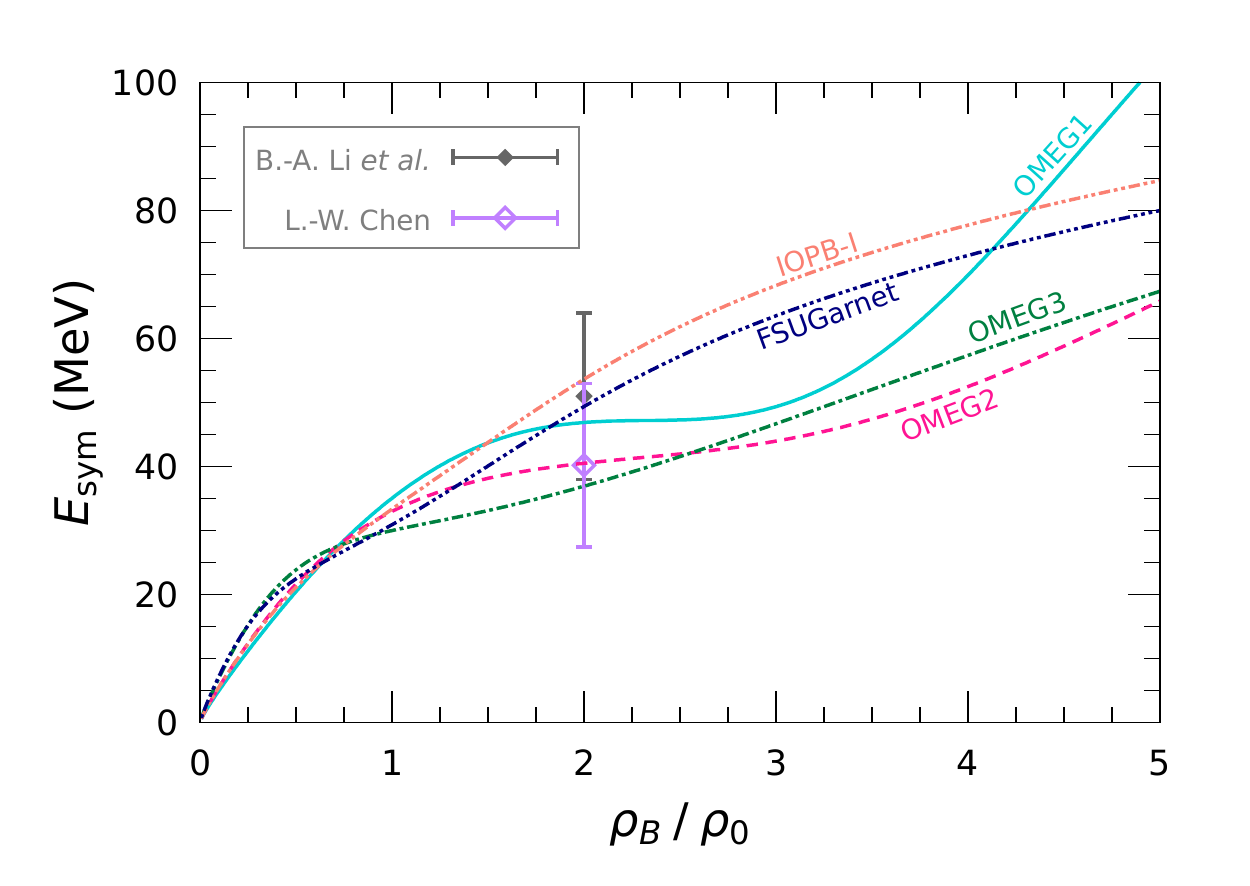}%
  \caption{\label{fig:Esym}
    Nuclear symmetry energy, $E_{\rm sym}$, as a function of $\rho_{B}/\rho_{0}$.}
\end{figure}
The density dependence of nuclear symmetry energy, $E_{\rm sym}$, is shown in Fig.~\ref{fig:Esym}.
We also refer to the theoretical constraints on the magnitude of $E_{\rm sym}$ at $2\rho_{0}$: $E_{\rm sym}(2\rho_{0})\simeq40.2\pm12.8$~MeV based on microscopic calculations with various energy density functionals~\cite{Chen:2015gba}, and $E_{\rm sym}(2\rho_{0})\simeq51\pm13$~MeV from nine new analyses of neutron-star observables since GW170817~\cite{Li:2021thg}.
As already explained in our previous study~\cite{Miyatsu:2022wuy}, the $\sigma$--$\delta$ mixing suppresses $E_{\rm sym}$ above $2\rho_{0}$, which is similar to the tendency shown in the relativistic Brueckner-Hartree-Fock (RBHF) calculation~\cite{Katayama:2013zya,Wang:2022fqt} and the skyrmion crystal approach~\cite{Ma:2021nuf,Lee:2021hrw}.
This suppression of $E_{\rm sym}$ causes the softness of the EoS for pure neutron matter around $2\rho_{0}$, as mentioned in Fig.~\ref{fig:Pressure}(b).

\begin{figure}[t!]
  \includegraphics[width=8.5cm,keepaspectratio,clip]{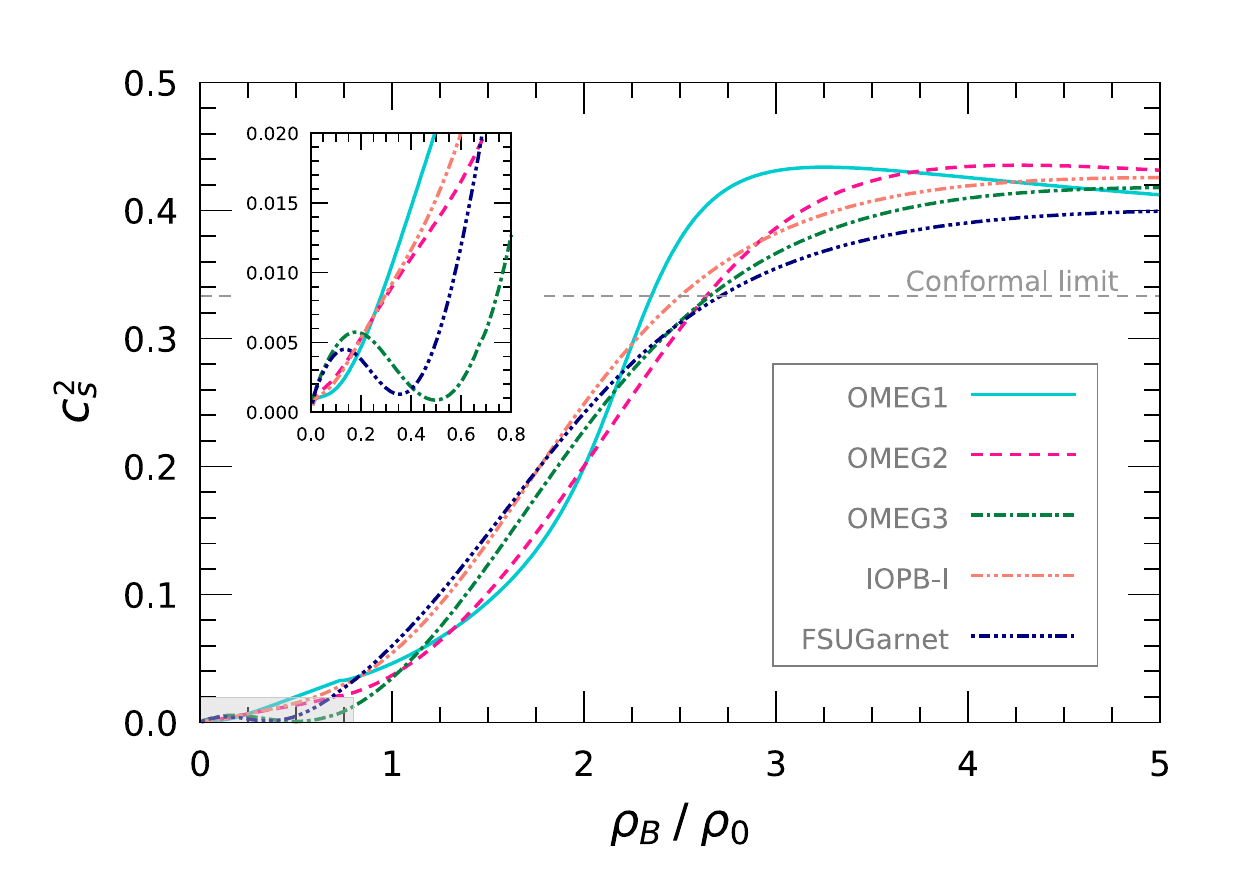}%
  \caption{\label{fig:Sound}
    Speed of sound squared, $c_{s}^{2}$, as a function of $\rho_{B}/\rho_{0}$, in units of the speed of light squared.
    The conformal limit is given by $c_{s}^{2}=1/3$~\cite{Alford:2013aca,Reed:2019ezm}.
    The inset displays $c_{s}^{2}$ at densities below $0.8\rho_{0}$.}
\end{figure}
In Fig.~\ref{fig:Sound}, we show the speed of sound squared, $c_{s}^{2}$, in neutron-star matter, where the charge neutrality and $\beta$ equilibrium under weak processes are imposed.
As explained by \citet{Alford:2013aca}, $c_{s}^{2}$ for all the cases lies above the conformal limit, $c_{s}^{2}=1/3$, at high densities.
On the other hand, the quick reduction of $c_{s}^{2}$ occurs at low densities in the OMEG3 and FSUGarnet, which relates to the small $R_{\rm skin}^{48}$ and $R_{\rm skin}^{208}$, as shown in Fig.~\ref{fig:Rskin}.
If we suppose the smaller $R_{\rm skin}$ and hence smaller $L$ than that for the OMEG3, then thermodynamic stability may be violated by negative incompressibility, which may lead to the multilayer structure of neutron stars~\cite{Kubis:2020ysv}.
Thus, it may be favorable that the value of $L$ is not lower than that for the OMEG3, namely $L\gtrsim20$ MeV.

In order to discuss the properties of neutron stars, it is essential to consider the low-density EoS which covers the crust region of a neutron star.
In the present study, we employ the realistic EoS for nonuniform matter~\cite{Miyatsu:2013hea,Miyatsu:2015kwa}, in which neutron-rich nuclei and neutron drips out of the nuclei are considered using the Thomas-Fermi calculation with the uniform nuclear EoS based on the relativistic Hartree-Fock (RHF) approximation.

\begin{figure}[t!]
  \includegraphics[width=8.5cm,keepaspectratio,clip]{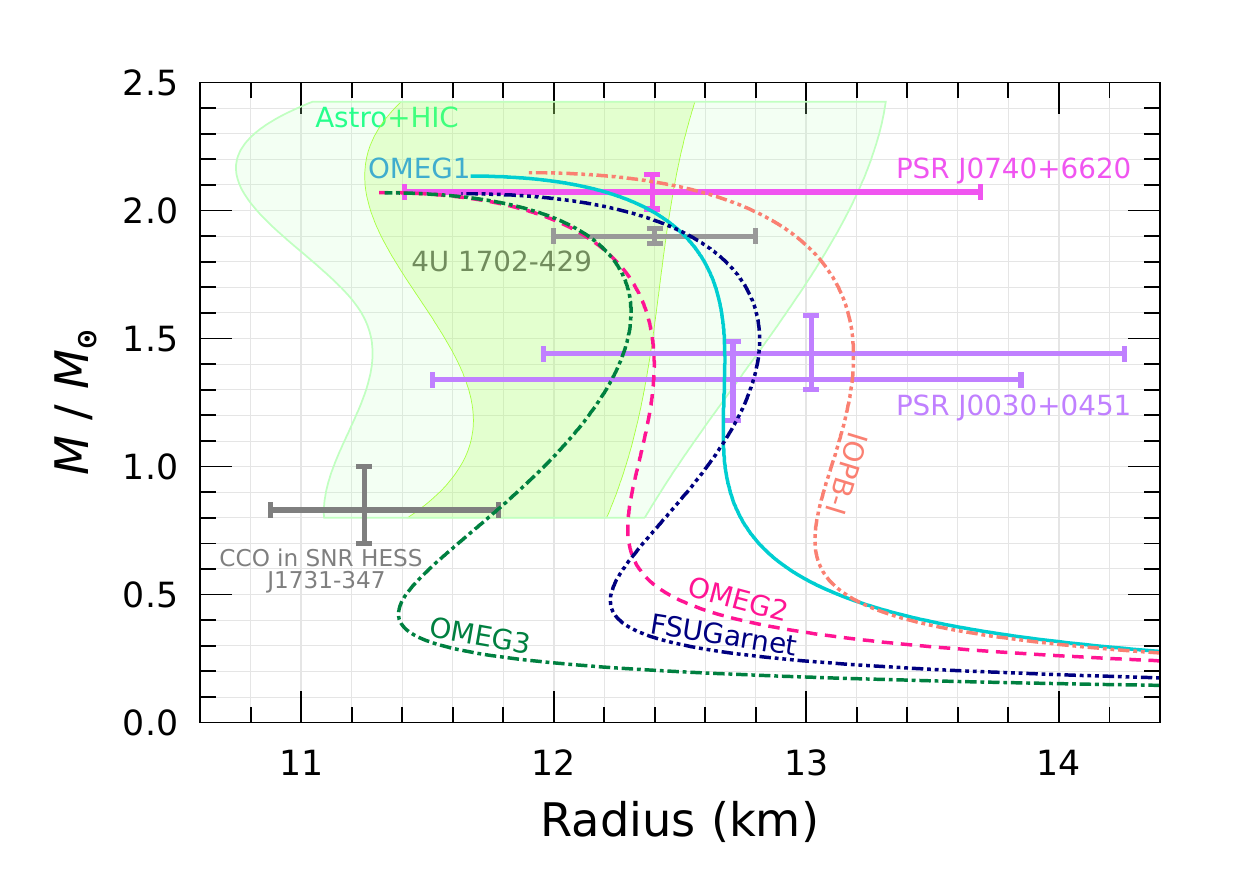}%
  \caption{\label{fig:MR}
    Mass--radius relations of neutron stars.}
\end{figure}
The mass--radius relations of neutron stars are then presented in Fig.~\ref{fig:MR}.
The observation data are here provided by the constraints from PSR J0030$+$0451 ($1.44_{-0.14}^{+0.15}$~$M_{\odot}$ and $13.02_{-1.06}^{+1.24}$~km, and $1.34_{-0.16}^{+0.15}$~$M_{\odot}$ and $12.71_{-1.19}^{+1.14}$~km)~\cite{Miller:2019cac,Riley:2019yda}, 4U 1702$-$429 ($1.9\pm0.3$~$M_{\odot}$ and $12.4\pm0.4$~km)~\cite{Nattila:2017wtj}, and PSR J0740$+$6620 ($2.072^{+0.067}_{-0.066}$~$M_{\odot}$ and $12.39^{+1.30}_{-0.98}$~km)~\cite{NANOGrav:2019jur,Fonseca:2021wxt,Riley:2021pdl}.
The shaded area shows the theoretical restriction using Bayesian inference based on the combined data from astrophysical multi-messenger observations of neutron stars and from HICs of gold nuclei at relativistic energies with microscopic calculations (Astro$+$HIC)~\cite{Huth:2021bsp}.
We also give the original information on a central compact object (CCO) within the supernova remnant (SNR), HESS J1731$-$347 ($0.83_{-0.13}^{+0.17}$~$M_{\odot}$ and $11.25_{-0.37}^{+0.53}$~km)~\cite{2022NatAs...6.1444D}.
We then summarize, in Table~\ref{tab:NS}, the properties of neutron stars at the canonical- and maximum-mass points, $M=1.4$~$M_{\odot}$ and $M=M_{\max}$.
\input{./Tables/Table-NS.tex}

We find that the OMEG family supports the astrophysical constraints on $M_{\max}$ from the observed mass of PSR J0740$+$6620 and on $\Lambda_{1.4}$ from GW170817 and GW190814.
In addition, their mass--radius relations adequately satisfy the observation data from 4U 1702$-$429 and PSR J0030$+$0451, and the theoretical restriction based on the combined data from multi-messenger observations and HICs (Astro$+$HIC).

We here emphasize that, compared with the IOPB-I and FSUGarnet, the OMEG family produces the smaller neutron-star radius, $R_{1.4}$, and the larger central density, $\rho_{1.4}$, at $M=1.4$~$M_{\odot}$ due to the $\sigma$--$\delta$ mixing.
They can thus afford to explain the severe constraint on $\Lambda_{1.4}=190^{+390}_{-120}$ from GW170817~\cite{LIGOScientific:2018cki}.
Especially, the OMEG1 provides the largest $R_{1.4}$ in the OMEG family because of the large $L=70$ MeV given through the calibration of $R_{\rm skin}^{208}$ by the PREX-2 result.
If $L$ is supposed to be larger than that for the OMEG1, it may be impossible to preserve $\Lambda_{1.4}$ from GW170817.
In contrast, the OMEG3, which can describe most experimental data of $R_{\rm skin}^{48}$ and $R_{\rm skin}^{208}$ (except for the PREX-2 result) with $L=20$~MeV, gives the smallest $R_{1.4}$ in all the models.
We can also point out the possibility that the CCO within the SNR, HESS J1731$-$347, might be the lightest neutron star if the EoS in the OMEG3 is applied.

\section{Summary and conclusion} \label{sec:summary}

We have constructed new effective interactions, labeled as OMEG1, OMEG2, and OMEG3, using the RMF model with nonlinear couplings between the isoscalar and isovector mesons.
By introducing the $\delta$--$N$ coupling and $\sigma$--$\delta$ mixing, we have studied the ground-state properties of finite, closed-shell nuclei as well as the characteristics of nuclear and neutron-star matter.
Considering the constraints from the particle flow data in HICs~\cite{Danielewicz:2002pu,Fuchs:2005zg,Lynch:2009vc}, the observed mass of PSR J0740$+$6620~\cite{NANOGrav:2019jur,Fonseca:2021wxt,Riley:2021pdl}, and $\Lambda_{1.4}$ from binary merger events, GW170817 and GW190814~\cite{LIGOScientific:2018cki,LIGOScientific:2020zkf}, as well as the PREX-2 and CREX results~\cite{PREX:2021umo,CREX:2022kgg}, we have calibrated the model parameters so as to reproduce the experimental data for $B/A$ and $R_{\rm ch}$ of several closed-shell nuclei.

It has been found that the $\delta$--$N$ coupling and $\sigma$--$\delta$ mixing much affect the properties of isospin-asymmetric nuclear matter and finite nuclei, and that they play an important role in understanding of both terrestrial experiments and astrophysical observations of neutron stars.
In particular, the OMEG family can satisfy the stringent constraint on $\Lambda_{1.4}=190^{+390}_{-120}$ from GW170817 because the $\sigma$--$\delta$ mixing suppresses $E_{\rm sym}$ above $2\rho_{0}$.
Taking into account the PREX-2 data~\cite{PREX:2021umo}, we have obtained $R_{\rm skin}^{208}=0.243$~fm using $L=70$~MeV in the OMEG1.
In contrast, the CREX data~\cite{CREX:2022kgg} as well as the experimental data of $^{48}$Ca and $^{208}$Pb at RCNP~\cite{Birkhan:2016qkr,Tamii:2011pv} and of $^{208}$Pb at MAMI ~\cite{Tarbert:2013jze} suggests the small $L$, for example, as in the OMEG3.
We have thus concluded that the PREX-2 and CREX results cannot be reconciled with each other within the present RMF model including isoscalar-isovector couplings with the constraints from various high density data.

Finally, we note that it is interesting to study the role of isovector self-interactions, $(\bm{\delta}\cdot\bm{\delta})^{2}$ and $(\bm{\rho}_{\mu}\cdot\bm{\rho}^{\mu})^{2}$, on the properties of finite nuclei and the EoS for neutron stars in the RMF and relativistic Hartree-Fock calculations~\cite{Miyatsu:2011bc,Katayama:2012ge,Miyatsu:2020vzi}.
It is also important to consider the isospin symmetry breaking effect on asymmetric nuclear matter from the quark level inside a nucleon~\cite{Guichon:1987jp,Saito:1994ki,Saito:1994tq,Saito:2005rv,Nagai:2008ai}.

\section*{Acknowledgments}

This work was supported by the National Research Foundation of Korea (Grant Nos. NRF-2021R1A6A1A03043957, NRF-2020R1A2C3006177, and 2018R1A5A1025563).

\bibliographystyle{elsarticle-num-names-mod.bst}
\bibliography{RMF-OMEG.bib}

\end{document}

%% file: Tables/Table-CCs.tex
\begin{table*}[t!]
  \centering
  \caption{\label{tab:CCs}
    Model parameters for the OMEG family, IOPB-I, and FSUGarnet.
    The $\delta$-meson mass in free space is fixed at $m_{\delta}=980.000$~MeV.}
  \resizebox{\textwidth}{!}{
    \begin{tabular}{lrrrrrrrrrrrr}
      \toprule
      \         & $m_{\sigma}$ & $m_{\omega}$ & $m_{\rho}$ & $g_{\sigma}^{2}$ & $g_{\omega}^{2}$ & $g_{\delta}^{2}$ & $g_{\rho}^{2}$ & $g_{2}$     &   $g_{3}$ & $c_{3}$ & $\Lambda_{\sigma\delta}$ & $\Lambda_{\omega\rho}$ \\
      Models    &        (MeV) &        (MeV) &      (MeV) &                \ &                 \ &               \ &              \ & (fm$^{-1}$) &         \ &      \  &                        \ &                      \ \\
      \midrule
      OMEG1     &      497.825 &      782.660 &    775.260 &           99.645 &          166.268 &           30.000 &         44.591 &       7.824 &  $-1.115$ & 100.000 &                   95.000 &                 75.677 \\
      OMEG2     &      497.820 &      782.660 &    775.260 &           99.641 &          166.269 &           20.000 &         44.364 &       7.823 &  $-1.113$ & 100.000 &                   85.000 &                288.859 \\
      OMEG3     &      498.015 &      782.660 &    775.260 &           99.713 &          166.272 &           15.000 &         57.550 &       7.827 &  $-1.105$ & 100.000 &                   70.000 &                909.825 \\
      IOPB-I    &      500.487 &      782.500 &    762.500 &          108.012 &          177.981 &              --- &         30.906 &      10.502 & $-14.802$ &  92.108 &                      --- &                527.225 \\
      FSUGarnet &      496.939 &      782.500 &    763.000 &          110.350 &          187.693 &              --- &         47.966 &       9.565 &  $-7.122$ & 137.981 &                      --- &               1555.728 \\
      \bottomrule
    \end{tabular}
  }
\end{table*}

%% file: Tables/Table-finite.tex
\begin{table*}[t!]
  \centering
  \caption{\label{tab:finite}
    Theoretical predictions for ground-state properties of several closed-shell nuclei.
    Experimental data are referred to Refs.~\cite{Angeli:2013epw,Wang:2021xhn,Malbrunot-Ettenauer:2021fnr}.}
  \begin{tabular}{lcccccccccc}
    \toprule
    \           & \multicolumn{9}{c}{$B/A$ (MeV)} \\
    \cline{2-10}
    Models      &  $^{16}$O & $^{40}$Ca & $^{48}$Ca & $^{68}$Ni & $^{90}$Zr & $^{100}$Sn & $^{116}$Sn & $^{132}$Sn & $^{208}$Pb & $\Delta_{\rm RRSE}$ ($\%$) \\
    \midrule
    OMEG1       &      7.99 &      8.55 &      8.60 &      8.67 &      8.70 &       8.26 &       8.51 &       8.35 &       7.90 &               0.31 \\
    OMEG2       &      7.99 &      8.55 &      8.59 &      8.67 &      8.69 &       8.26 &       8.51 &       8.35 &       7.91 &               0.38 \\
    OMEG3       &      7.99 &      8.55 &      8.57 &      8.67 &      8.69 &       8.26 &       8.52 &       8.35 &       7.92 &               0.46 \\
    IOPB-I      &      8.11 &      8.59 &      8.65 &      8.70 &      8.69 &       8.27 &       8.49 &       8.35 &       7.87 &               0.63 \\
    FSUGarnet   &      8.01 &      8.53 &      8.61 &      8.68 &      8.69 &       8.28 &       8.49 &       8.36 &       7.90 &               0.34 \\
    \midrule
    Experiments &      7.98 &      8.55 &      8.67 &      8.68 &      8.71 &       8.25 &       8.52 &       8.35 &       7.87 &                --- \\
    \midrule
    \           & \multicolumn{9}{c}{$R_{\rm ch}$ (fm)} \\
    \cline{2-10}
    Models      &  $^{16}$O & $^{40}$Ca & $^{48}$Ca & $^{68}$Ni & $^{90}$Zr & $^{100}$Sn & $^{116}$Sn & $^{132}$Sn & $^{208}$Pb & $\Delta_{\rm RRSE}$ ($\%$) \\
    \midrule
    OMEG1       &      2.74 &      3.48 &      3.50 &      3.89 &      4.29 &       4.50 &       4.63 &       4.73 &       5.53 &               0.67 \\
    OMEG2       &      2.74 &      3.48 &      3.50 &      3.90 &      4.29 &       4.50 &       4.63 &       4.74 &       5.54 &               0.71 \\
    OMEG3       &      2.74 &      3.48 &      3.51 &      3.91 &      4.30 &       4.50 &       4.65 &       4.75 &       5.55 &               0.88 \\
    IOPB-I      &      2.74 &      3.48 &      3.49 &      3.89 &      4.29 &       4.48 &       4.63 &       4.74 &       5.55 &               0.73 \\
    FSUGarnet   &      2.73 &      3.46 &      3.47 &      3.87 &      4.26 &       4.44 &       4.61 &       4.71 &       5.51 &               0.46 \\
    \midrule
    Experiments &      2.70 &      3.48 &      3.48 &      3.89 &      4.27 &        --- &       4.63 &       4.71 &       5.50 &                --- \\
    \bottomrule
  \end{tabular}
\end{table*}

%% file: Tables/Table-matter.tex
\begin{table*}[t!]
  \centering
  \caption{\label{tab:matter}
    Properties of symmetric nuclear matter at $\rho_{0}$.
    Here, $E_{0}$ denotes the binding energy per nucleon.
    The bulk properties are given by coefficients in the power-series expansion of isospin-asymmetric nuclear EoS around $\rho_{0}$~\cite{Chen:2009wv}.
    The $K_{\rm sat,2}$ and $K_{N}$ are respectively expressed as $K_{\rm sat,2}(=K_{\tau})=K_{\rm sym}-6L-J_{0}L/K_{0}$ and $K_{N}=K_{\rm sym}+K_{0}$~\cite{Sotani:2022zvu,Lattimer:2023rpe}.}
  \begin{tabular}{llrrrrrrrrrr}
    \toprule
    \         &  $\rho_{0}$ & $M_{N}^{\ast}$ &  $E_{0}$ & $K_{0}$ &     $J_{0}$ & $E_{\rm sym}$ &   $L$ & $K_{\rm sym}$ & $J_{\rm sym}$ & $K_{\rm sat,2}$ & $K_{N}$ \\
    Models    & (fm$^{-3}$) &      ($M_{N}$) &    (MeV) &   (MeV) &       (MeV) &         (MeV) & (MeV) &         (MeV) &         (MeV) &           (MeV) &   (MeV) \\
    \midrule
    OMEG1     &      0.1484 &          0.620 & $-16.38$ &  256.00 &   $-300.62$ &         35.06 & 70.00 &     $-218.83$ &      $-68.93$ &       $-556.62$ &   37.17 \\
    OMEG2     &      0.1484 &          0.620 & $-16.38$ &  256.00 &   $-300.56$ &         33.00 & 45.00 &     $-216.72$ &       1020.63 &       $-433.89$ &   39.28 \\
    OMEG3     &      0.1484 &          0.620 & $-16.38$ &  256.00 &   $-300.28$ &         30.00 & 20.00 &      $-65.85$ &       1450.00 &       $-162.39$ &  190.15 \\
    IOPB-I    &      0.149  &          0.595 & $-16.11$ &  222.65 &   $-116.44$ &         33.36 & 63.70 &      $-40.13$ &        873.02 &       $-389.00$ &  182.53 \\
    FSUGarnet &      0.153  &          0.579 & $-16.23$ &  229.63 &        8.47 &         30.92 & 50.96 &         58.32 &        149.11 &       $-249.33$ &  287.94 \\
    \bottomrule
  \end{tabular}
\end{table*}

%% file: Tables/Table-NS.tex
\begin{table}[t!]
  \centering
  \caption{\label{tab:NS}
    Properties of neutron stars at the canonical- and maximum-mass points.}
  \resizebox{\hsize}{!}{
    \begin{tabular}{lcccccc}
      \toprule
      \         & $R_{1.4}$ & $\rho_{1.4}$ & $\Lambda_{1.4}$ &    $M_{\max}$ & $R_{\max}$ & $\rho_{\max}$\\
      Models    &      (km) & (fm$^{-3}$)  &               \ & ($M_{\odot}$) &       (km) &  (fm$^{-3}$) \\
      \midrule
      OMEG1     &     12.68 &         0.40 &             529 &          2.13 &      11.67 &         0.92 \\
      OMEG2     &     12.40 &         0.43 &             471 &          2.07 &      11.31 &         0.98 \\
      OMEG3     &     12.26 &         0.41 &             512 &          2.07 &      11.33 &         0.96 \\
      IOPB-I    &     13.19 &         0.36 &             699 &          2.15 &      11.90 &         0.89 \\
      FSUGarnet &     12.81 &         0.38 &             635 &          2.07 &      11.63 &         0.93 \\
      \bottomrule
    \end{tabular}
    }
\end{table}